\documentclass[twocolumn,showpacs,preprintnumbers,amsmath,amssymb,pra]{revtex4}
\usepackage{graphicx}
\usepackage{dcolumn}

\begin{document}
\title{Dependence of the evolution of the cavity radiation of a coherently pumped correlated emission laser on dephasing and phase fluctuation}

\author{Sintayehu Tesfa}
\affiliation{Max Planck Institute for the Physics of Complex Systems, N$\ddot{o}$thnitzer Str. 38, 01187 Dresden, Germany\\Physics Department, Dilla University, P. O. Box 419, Dilla, Ethiopia}

\date{\today}

\begin{abstract} Analysis of the dynamics of the cavity radiation of a coherently pumped correlated emission laser is presented. The phase fluctuation and dephasing are found to affect the time evolution of the two-mode squeezing and intensity of the cavity radiation significantly. The intensity and degree of the two-mode squeezing increase at early stages of the process with time, but this trend changes rapidly afterwards. It is also shown that they increase with phase fluctuation and dephasing in the strong driving limit, however the situation appears to be opposite in the weak driving limit. This essentially suggests that the phase fluctuation and dephasing weaken the coherence induced by a strong driving mechanism so that the spontaneous emission gets a chance. The other important aspect of the phase fluctuation, in this regard, is the relaxation of the time at which the maximum squeezing is manifested as well as the time in which the radiation remains in a squeezed state.\end{abstract}

 \maketitle

 \section{INTRODUCTION}

In a two-photon three-level laser, the  nonclassical features of the radiation are attributed to the atomic coherence  induced by preparing the atoms initially in a coherent superposition of the energy levels between which a direct spontaneous transition is electric dipole forbidden \cite{pra74043816,jpbamos,pra49481,prl601832} or by pumping the same with external radiation \cite{prl94023601,pra75033816,jpb41145501,pra415179,pra461560,pra72022305}. The roles of these mechanisms on the generation of light with nonclassical features have been extensively investigated elsewhere \cite{pra74043816,jpbamos,pra49481,prl601832,prl94023601,pra75033816,jpb41145501,pra415179,pra461560,pra72022305,pra484686}. The earlier studies unequivocally assert that the various schemes of this laser can be a potential candidate for the realization of a strong bright entangled light. However, a number of naturally occurring phenomena that significantly alter the degree of nonclassicality were ignored so far for various reasons. 

To begin with, as recently discussed, the present practical capability cannot provide a foolproof for a perfect preparation of the atomic coherent superposition \cite{sint,pra79013831,oc283781,pra444688}. In other words, the actually induceable coherent superposition is less than what is assumed due to loses arising from faulty setups. Even though there is a proposal acclaiming that the phase fluctuations that emerge in connection with the faulty preparation might enhance nonclassicality via creating indirect spontaneous emission roots \cite{sint}, a detailed analysis is still missing to attest the claim. The other noteworthy mechanism that has a potential for destroying the coherence is the dephasing that originates from loses of atomic coherence through various quantum processes. In light of this, in a case when it is possible to lock the phase, it has been argued that there is a chance for enhancing the nonclassicality \cite{pra79063815}. When there is no external intervention, it is more likely that the atomic coherence decays faster than the atomic decay rates.

Moreover, the study of the time development of the two-mode squeezing and entanglement has received a considerable attention in recent years \cite{pra72022305,pra79013831,pra77062308,pra444688,oc283781}.  It has been generally observed that the two-mode squeezing and entanglement increase at early stages of the process and later found to decrease with time \cite{oc283781,pra444688,pra72022305,pra79013831}. Particularly, in the pumped correlated emission laser when the atoms are initially prepared to be perfectly in the lower energy level and the external pumping radiation induces phase fluctuations, the emerging phase fluctuations turn out to decrease the entanglement for all cases \cite{pra77062308}. Besides, the effect of the phase fluctuations is found to be prominent when the cavity is assumed to contain photons initially \cite{pra79013831}. Investigation of a similar nature with no external driving radiation when the atoms are partially prepared indicates that the phase fluctuations induced due to imperfect preparation decrease entanglement and the time for which the system remains entangled \cite{oc283781}. 

With this background, it appears natural to raise some basic questions. Is there a chance for the phase fluctuations to enhance the entanglement? Assuming that it is technically possible, what happens if the atoms are externally driven by a strong coherent light? Can the degree of entanglement increase if the coherent superposition decays with larger rate than the spontaneous atomic decay rate? The fundamental issue behind these questions is the suspicion emanating from the competing nature of the various phenomena in relation to the coherence responsible for the manifestation of the nonclassical correlations. To answer these and other similar questions, a comprehensive approach in which these competing effects are brought together is required. 

Hence, in this contribution, coherently pumped correlated emission laser in which the three-level atoms are initially prepared partially in 50:50 probability to be in the upper and lower energy levels is considered (detailed description of the physical model can be found for instance in \cite{jpb41055503}). It is not difficult to envision at this juncture that the pumping radiation has associated phase fluctuations due to instability in the laser. It has been reported elsewhere that the effect of such fluctuations on the generated entanglement is quite perceivable specially when there is initial radiation \cite{pra75062305,pra77062308,pra79013831}. Despite this evidence, the phase fluctuations related to the imperfect preparation is considered here. 

In addition, in connection with recent observation that the rate at which the coherent superposition decays contributes significantly towards the evolution of the entanglement \cite{pra75062305} and a proposal for the importance of including the rate of dephasing in the analysis \cite{pra79033810}, the rate at which the coherence superposition decays is assumed to be greater than the atomic decay rate. This assumption principally addresses the concern that the physical phenomena like vacuum fluctuations and broadening of the atomic levels can lead to loses of coherence without necessarily resulting atomic transitions. Taking these as motivation, the effect of the processes that induce coherence (initial preparation and external pumping) in one hand and the processes that destroy coherence (dephasing and phase fluctuation) on the other hand on the evolution of the two-mode squeezing and intensity of the cavity radiation is investigated. 

In the study of the time evolution of similar system when many parameters are involved, the usual approach is numerically calculating the required correlations using either the characteristic function \cite{pra79013831,pra72022305} or the rate equation \cite{pra77062308,oc283781}. However, in this paper, the dynamics of the cavity radiation is analyzed following a straightforward analytic approach. In the first place, the rate equations corresponding to the expectation values of the radiation modes ($\langle\hat{a}\rangle$ and $\langle\hat{b}\rangle$) are derived applying the master equation. Then the resulting expressions are converted into equations of c-number variables associated with the normal ordering. After that the obtained coupled differential equations are analytically solved employing the procedure outlined in \cite{pra79033810} and reference there in. Once the solutions are obtained various correlation are determined with the aid of the correlations of the pertinent Langevin noise forces. 

\section{Equations of evolution}

Interaction of a pumped nondegenerate three-level cascade atom with a resonant two-mode cavity
radiation can be described in the rotating-wave approximation and
the interaction picture by the Hamiltonian of the form
\begin{align}\label{p01}\hat{H} &=ig[\hat{a}|a\rangle\langle
b| - |b\rangle\langle a|\hat{a}^{\dagger} +\hat{b}|b\rangle\langle
c| - |c\rangle\langle b|\hat{b}^{\dagger}]\notag\\& +
i{\Omega\over2}[|c\rangle\langle a|-|a\rangle\langle c|],\end{align}
where $\Omega$ is a real-positive constant proportional to the
amplitude of the driving radiation and $g$ is a coupling constant
chosen to be the same for both transitions. $\hat{a}$ and $\hat{b}$
are the annihilation operators that represent the two cavity modes. In the cascade configuration,
the transition from upper energy level $|a\rangle$ to the intermediate
level $|b\rangle$ and from level $|b\rangle$ to the lower energy level
$|c\rangle$ are taken to be resonant with the cavity radiation,
whereas the transition $|a\rangle\leftrightarrow|c\rangle$ is electric dipole forbidden.
Although the pumping laser undoubtedly has a bandwidth that consequently leads to phase fluctuations, this issue is not considered  here to limit the involved  rigor. However, due to emerging various quantum effects, it is assumed that the atoms can only be initially prepared in a partial maximum coherent superposition of the
upper and lower energy levels. 

In view of these, the initial state of the three-level atom is taken to be
\begin{align}\label{p02}|\Psi_{A}(0)\rangle\; ={1\over\sqrt{2}}\big[|a\rangle +
e^{i\varphi}|c\rangle\big],\end{align} where  $\varphi$ is an arbitrary phase randomly distributed about a fixed mean value $\varphi_{0}$. As a result, the phase can be defined as $\varphi=\varphi_{0}+\delta\varphi$ where $\delta\varphi$ is taken to be small random fluctuations around $\varphi_{0}$ which can be adjusted at will by proper choice of the phase of the cavity radiation \cite{oc283781}. Hence one can set $\varphi_{0}=0$ and take $\varphi$ as fluctuations about central value 0. In line with Eq. \eqref{p02}, the
corresponding initial reduced atomic density operator appears to be
\begin{align}\label{p03}\hat{\rho}_{A}(0)& =
{1\over2}\big[|a\rangle\langle a| + e^{-i\varphi}|a\rangle\langle
c| + e^{i\varphi}|c\rangle\langle a| +
|c\rangle\langle c|\big],\end{align} where
$e^{\pm i\varphi}/2$ stand for the initial atomic coherence. 

In setting up the laser, the atoms prepared in this manner are assumed to be injected into the cavity at a constant rate $r_{a}$ and removed after sometime which is long enough for the atoms to decay spontaneously to energy levels that are not involved in the process. The rate at which the atoms decay through spontaneous emission $\Gamma$ is assumed to be less than the rate at which the atomic coherence superposition decays $\gamma$. Assuming a good cavity limit, the case in which the cavity damping rate $\kappa$ is much smaller than various channels of atomic decay rates ($\Gamma$ and $\gamma$), the adiabatic approximation scheme is employed. For the sake of convenience, these damping constants are taken to be the same for different modes and energy levels as the case may be. 

Moreover, the linear analysis is applied in which all terms are limited to the  square of the coupling constant $g$ so that the resulting differential equations can be analytically solvable. In the study of the nondegenerate three-level laser, the linearization procedure is usually applied in connection to the fact that the quantum features are by large associated with the coherence induced in the cascading process. With this background, applying the linear and adiabatic approximation schemes in the good cavity limit, the time evolution of the reduced density operator for the cavity radiation is found following the procedure outlined in \cite{pra79033810,sint} to be 
\begin{align}\label{p04}\frac{d\hat{\rho}}{dt}& = \frac{\kappa}{2}
[2\hat{a}\hat{\rho}\hat{a}^{\dagger} -
 \hat{a}^{\dagger}\hat{a}\hat{\rho} -
 \hat{\rho}\hat{a}^{\dagger}\hat{a}] \notag\\&+
\frac{AC}{2B} \left[2\hat{a}^{\dagger}\hat{\rho}\hat{a} -
 \hat{\rho}\hat{a}\hat{a}^{\dagger} -
 \hat{a}\hat{a}^{\dagger}\hat{\rho}\right]\notag\\&+
\frac{1}{2}\left(\frac{AC}{B}+\kappa\right)\left[2\hat{b}\hat{\rho}\hat{b}^{\dagger} -
\hat{\rho}\hat{b}^{\dagger}\hat{b} -
\hat{b}^{\dagger}\hat{b}\hat{\rho}\right]\notag\\&
+\frac{AD_{+}}{2B}\left[\hat{b}\hat{\rho}\hat{b}^{\dagger}-\hat{a}^{\dagger}\hat{\rho}\hat{a}-\hat{b}^{\dagger}\hat{b}\hat{\rho}
+\hat{a}\hat{a}^{\dagger}\hat{\rho}\right]\notag\\&
+\frac{AD_{-}}{2B}\left[\hat{b}\hat{\rho}\hat{b}^{\dagger}-\hat{a}^{\dagger}\hat{\rho}\hat{a}-\hat{\rho}\hat{b}^{\dagger}\hat{b}
+\hat{\rho}\hat{a}\hat{a}^{\dagger}\right]\notag\\&
-\frac{AE_{+}}{2B}\left[
\hat{a}^{\dagger}\hat{\rho}\hat{b}^{\dagger} -
\hat{b}^{\dagger}\hat{a}^{\dagger}\hat{\rho}+
\hat{b}\hat{\rho}\hat{a}-\hat{a}\hat{b}\hat{\rho}\right]\notag\\&-\frac{AE_{-}}{2B}\left[\hat{a}^{\dagger}\hat{\rho}\hat{b}^{\dagger}
- \hat{\rho}\hat{b}^{\dagger}\hat{a}^{\dagger}+
\hat{b}\hat{\rho}\hat{a}-\hat{\rho}\hat{a}\hat{b}\right]\notag\\&
+\frac{A\zeta'(1+\zeta'\zeta)}{2B}\left[\hat{b}^{\dagger}\hat{a}^{\dagger}\hat{\rho}-\hat{\rho}\hat{b}^{\dagger}\hat{a}^{\dagger}-\hat{a}\hat{b}\hat{\rho}+\hat{\rho}\hat{a}\hat{b}\right],\end{align} where 
\begin{align}\label{p05}A = \frac{2r_{a}g^{2}}{\gamma^{2}},\end{align}
\begin{align}\label{p06}B=(4+\zeta^{2})(1+\zeta'\zeta),\end{align}
\begin{align}\label{p07}C&=2(\zeta'^{2}+\chi),\end{align}
\begin{align}\label{p08}D_{\pm}&=(2\zeta'+\zeta)e^{\pm i\varphi},\end{align}
\begin{align}\label{p09}E_{\pm}&=(2-\zeta'\zeta)e^{\pm i\varphi},\end{align} in which $\zeta={\Omega/\gamma}$, $\zeta'={\Omega/\Gamma}$, and $\chi=\gamma/\Gamma.$ 
The contribution of the cavity damping which corresponds to the coupling of the cavity modes with the two-mode vacuum reservoir via the coupler mirror is incorporated following the usual standard approach \cite{lou,scully}. 

Comparison reveals that this master equation has a fundamental difference in form from the earlier report of similar scheme \cite{jpb41055503}. It looks from the outset that  the gain of mode $a$ and the lose of mode $b$ are equal. Even though the terms associated with $C$ have the same form with earlier reports where such deduction is possible, in Eq. \eqref{p04}, some of these terms are unambiguously included in $D_{\pm}$. As recently proposed,  such modification in the master equation emanates from the changes in the subsequent atomic transitions due to phase fluctuations introduced during preparation \cite{sint}. It is also vividly noticeable that the last term in the master equation closely related to the manifestation of the nonclassical features is solely attributed to changes of sign in $E_{\pm}$ due to phase fluctuations. Moreover, critical observation of the master equation reveals that the sign in front of $\hat{b}^{\dagger}\hat{a}^{\dagger}\hat{\rho}$ and $\hat{\rho}\hat{a}\hat{b}$ in $E_{\pm}$ and the last term is different. This entails that the external driving and partial initial preparation mechanisms have competing effect in which the quantum features that is lost due to faulty preparation setups can be compensated by adjusting the amplitude of the external driving radiation. 

On the other hand, as clearly demonstrated elsewhere \cite{pra74043816,pra79033810,sint}, the stochastic differential equations associated
with the normal ordering of the cavity mode variables are found making use of Eq. \eqref{p04} to be
\begin{align}\label{p10}\frac{d}{dt}\alpha(t) &= -\frac{B\kappa+A[(2\zeta'+\zeta)e^{-\theta}-2(\zeta'+\chi)]}{2B}\alpha(t)\notag\\&+ \frac{A\left[\zeta'(1+\zeta\zeta')-(2-\zeta'\zeta)e^{-\theta}\right]}{2B}\beta^{*}(t)+f_{a}(t),\end{align}
\begin{align}\label{p11}\frac{d}{dt}\beta(t)& =
-\frac{B\kappa+A[(2\zeta'+\zeta)e^{-\theta}+2(\zeta'+\chi)]}{2B}\beta(t)\notag\\& + \frac{A\left[\zeta'(1+\zeta\zeta')+(2-\zeta'\zeta)e^{-\theta}\right]}{2B}\alpha^{*}(t)+f_{b}(t),\end{align}
where $f_{a}(t)$ and $f_{b}(t)$ are the stochastic noise forces whose various correlations turn out to be
\begin{align}\label{p20}\langle f_{a}(t')f^{*}_{a}(t)\rangle & = {A\over B}\left[2\zeta'^{2}+2\chi-(2\zeta'+\zeta)e^{-\theta}\right] \delta(t-t'),\end{align}
\begin{align}\label{p21}\langle f_{b}(t')f^{*}_{b}(t)\rangle = 0,\end{align}
\begin{align}\label{p22}\langle f_{b}(t')f_{a}(t)\rangle&=
\frac{A}{2B}\left[\zeta'(1+\zeta'\zeta)+(2-\zeta'\zeta)e^{-\theta}\right]\delta(t-t'),\end{align}
\begin{align}\label{p23}\langle f^{*}_{b}(t')f_{a}(t)\rangle=\langle f_{a}(t')f_{a}(t)\rangle
=\langle f_{b}(t')f_{b}(t)\rangle = 0.\end{align}
While deriving the above equations, a stochastic average is taken over the phase since it is assumed to fluctuate about the central mean value (set to 0 for convenience). From practical point of view, addressing the contribution of every phase change appears to be unrealistic. In connection to this, assuming that the phase undergoes Gaussian random process \cite{pra77062308,pra444688,oc283781}, it would be more appropriate if the phase fluctuation is used instead of the actual phase. With this understanding and employing the fact that for Gaussian variables 
\begin{align}\label{p15}\langle\exp\pm i\delta\varphi\rangle=\exp-\langle\delta\varphi^{2}/2\rangle\end{align}   $\exp\pm i\varphi$  is replaced by $\exp-\theta$  \cite{method}. It is not difficult to realize that for Gaussian random process $\langle\delta\phi\rangle$ is zero, hence $\theta=\langle\delta^{2}\phi/2\rangle$ represents the deviation which is generally designated as phase fluctuation. 

In order to see the time evolution of each variable, these coupled differential equations are solved following the approach outlined in \cite{pra79033810,sint}. The solutions are
\begin{align}\label{p24}\alpha(t) = F_{+}(t)\alpha(0) + G_{+}(t)\beta^{*}(0) + H_{+}(t)
 + I_{+}(t),\end{align}
\begin{align}\label{p25}\beta(t) =  F_{-}(t)\beta(0)+G_{-}(t)\alpha^{*}(0)
+ H_{-}(t) + I_{-}(t),\end{align}
 where
\begin{align}\label{p26}F_{\pm}(t) = \frac{1}{2}\left[(1\pm p)e^{-\mu_{-}t} +
(1\mp p)e^{-\mu_{+}t}\right],\end{align}
\begin{align}\label{p27}G_{\pm}(t) = \frac{q_{\pm}}{2}[e^{-\mu_{+}t} - e^{-\mu_{-}t}],\end{align}
\begin{align}\label{p28}H_{+}(t)& =  \frac{1}{2}\int_{0}^{t}[(1+p)e^{-\mu_{-}(t-t')} \notag\\&+
(1-p)e^{-\mu_{+}(t-t')}]f_{a}(t')dt',\end{align}
\begin{align}\label{p29}H_{-}(t) &= \frac{1}{2}\int_{0}^{t}[(1-p)e^{-\mu_{-}(t-t')}
\notag\\&+ (1+p)e^{-\mu_{+}(t-t')}]f_{b}(t')dt',\end{align}
\begin{align}\label{p30}I_{+}(t) = \frac{q_{+}}{2}\int_{0}^{t}[e^{-\mu_{+}(t-t')} -
e^{-\mu_{-}(t-t')}]f^{*}_{b}(t')dt',\end{align}
\begin{align}\label{p31}I_{-}(t) = \frac{q_{-}}{2}\int_{0}^{t}[e^{-\mu_{+}(t-t')} -
e^{-\mu_{-}(t-t')}]f_{a}^{*}(t')dt',\end{align} with
\begin{align}\label{p32}\mu_{\pm} &= {\kappa\over2}+{A\over2B}\left[(2\zeta'+\zeta)e^{-\theta}\right.\notag\\&\left.\pm\sqrt{\zeta'^{2}(1+\zeta\zeta')^{2}+4(\zeta'^{2}+\chi)^{2}-
[(2-\zeta'\zeta)e^{-\theta}]^{2}}\right],\end{align}
\begin{align}\label{p33}p={2\big[\zeta'^{2}+\chi\big]\over
\sqrt{\zeta'^{2}(1+\zeta\zeta')^{2}+4(\zeta'^{2}+\chi)^{2}-
[(2-\zeta'\zeta)e^{-\theta}]^{2}}},\end{align}
\begin{align}\label{p34}q_{\pm}={-\zeta'(1+\zeta'\zeta)\pm\big[(2-\zeta'\zeta)e^{-\theta}\big]
\over\sqrt{\zeta'^{2}(1+\zeta\zeta')^{2}+4(\zeta'^{2}+\chi)^{2}-
[(2-\zeta'\zeta)e^{-\theta}]^{2}}}.\end{align}
It perhaps worth mentioning that Eqs.  \eqref{p24} and \eqref{p25} along with the associated parameters
are used to calculate various quantities of interest.  

\section{Two-mode quadrature squeezing}

The two-mode squeezing for coherently pumped correlated emission laser has been studied at steady state when the atoms are assumed to be perfectly prepared initially \cite{jpb41055503}. Taking all atomic decay rates as equal, it has been reported that the two-mode squeezing exists either for smaller or larger amplitudes of the driving radiation. In connection to the restriction associated to steady state solutions, the analysis has been limited to smaller values of the linear gain coefficient. Comparison shows that the degree of squeezing is better for $\Omega=0.5\gamma$. In this section, following a similar approach, the evolution of the two-mode quadrature squeezing is evaluated and comparison with the steady state case is made whenever possible. In addition, the effects of phase fluctuation and dephasing on the degree of squeezing in relation to the pumping mechanism is investigated. The cases for which these phenomena, believed to degrade nonclassical features, aid in enhancing the squeezing is also looked for. 

Generally, a two-mode cavity radiation can be described by an
operator
\begin{align}\label{p35}\hat{c}={1\over\sqrt{2}}\big(\hat{a}+\hat{b}\big),\end{align}
 where $\hat{a}$ and $\hat{b}$ represent the separate modes. 
With this consideration, the squeezing properties of the
cavity radiation can be studied applying the quadrature operators
defined by
\begin{align}\label{p37}\hat{c}_{+}=\hat{c}^{\dagger}+\hat{c}\end{align}
and
\begin{align}\label{p38}\hat{c}_{-}=i(\hat{c}^{\dagger}-\hat{c}),\end{align} where the corresponding variance can be readily obtained in terms of c-number variables associated with the normal ordering as
\begin{align}\label{p43}\Delta c_{\pm}^{2}& =1+ \frac{1}{2}\big[2\langle\alpha^{*}\alpha\rangle
\pm\langle\alpha^{*^{2}}\rangle \pm \langle\alpha^{2}\rangle   -
\langle\alpha^{*}\rangle^{2} - \langle\alpha\rangle^{2} \notag\\&\pm
\langle\beta^{*^{2}}\rangle \pm \langle\beta^{2}\rangle +
2\langle\beta^{*}\beta\rangle - \langle\beta^{*}\rangle^{2} -
\langle\beta\rangle^{2}\big] \pm \langle\alpha^{*}\beta^{*}\rangle\notag\\&
+ \langle\alpha^{*}\beta\rangle + \langle\alpha\beta^{*}\rangle
\pm \langle\alpha\beta\rangle \mp
\langle\alpha^{*}\rangle\langle\alpha\rangle-
\langle\alpha^{*}\rangle\langle\beta^{*}\rangle \notag\\&\mp
\langle\alpha^{*}\rangle\langle\beta\rangle \mp
\langle\alpha\rangle\langle\beta^{*}\rangle -
\langle\alpha\rangle\langle\beta\rangle \mp
\langle\beta^{*}\rangle\langle\beta\rangle. \end{align}

The various correlations in Eq. \eqref{p43} can be
determined using Eqs. \eqref{p24} and \eqref{p25}. In this line, assuming the
cavity to be initially in a two-mode vacuum state and the noise force at time $t$ does not
statistically related to the cavity mode variables at earlier
times, one can
readily verify that
\begin{align}\label{p44}\langle\alpha(t)\rangle =
 \langle\beta(t)\rangle = 0,\end{align} 
\begin{align}\label{p45}\langle\alpha^{*}(t)\alpha(t)\rangle &=
A\left[\frac{L(1-p)^{2}+
Mq_{+}(1-p)}{8B\mu_{+}}\right][1 -
e^{-2\mu_{+}t}] \notag\\&+A\left[\frac{L(1+p)^{2}- Mq_{+}(1+p)}{8B\mu_{-}}\right][1-e^{-2\mu_{-}t}]
\notag\\&+A\left[ \frac{L(1-p^{2}) +Mq_{+}p}{2B(\mu_{+}+\mu_{-})}\right][1
- e^{-(\mu_{+}+\mu_{-})t}],\end{align}
\begin{align}\label{p46}\langle\beta^{*}(t)\beta(t)\rangle &= A\left[\frac{Lq_{-}^{2}
+Mq_{-}(1+p)}{8B\mu_{+}}\right][1 - e^{-2\mu_{+}t}]
\notag\\&+A\left[\frac{Lq_{-}^{2} -Mq_{-}(1-p)}{8B\mu_{-}}\right][1-e^{-2\mu_{-}t}]
\notag\\&-A\left[ \frac{Lq_{-}^{2}+ Mq_{-}p}{2B(\mu_{+}+\mu_{-})}\right][1 -
e^{-(\mu_{+}+\mu_{-})t}],\end{align}
\begin{align}\label{p47}\langle\alpha(t)\beta(t)\rangle &= A\left[\frac{2Lq_{-}(1-p)+M(1-p^{2}+q_{-}q_{+})}{16B\mu_{+}}\right]\notag\\&\times[1
- e^{-2\mu_{+}t}] \notag\\&-A\left[\frac{2Lq_{-}(1+p)- M(1-p^{2}+q_{-}q_{+})}{16B\mu_{-}}\right]\notag\\&\times[1-e^{-2\mu_{-}t}]
\notag\\&+A\left[ \frac{2Lq_{-}p +M(1+p^{2}-q_{-}q_{+})}{4B(\mu_{+}+\mu_{-})}\right]\notag\\&\times[1
- e^{-(\mu_{+}+\mu_{-})t}],\end{align}
\begin{align}\label{p48}\langle\alpha^{2}(t)\rangle=\langle\beta^{2}(t)\rangle=\langle\alpha^{*}(t)\beta(t)\rangle=0,\end{align} where
\begin{align}\label{p49}L=2\zeta'^{2}+2\chi-(2\zeta'+\zeta)e^{-\theta},\end{align}
\begin{align}\label{p50}M=\zeta'(1+\zeta'\zeta)+(2-\zeta'\zeta)e^{-\theta}.\end{align}

Hence on account of Eqs. \eqref{p44} and \eqref{p48}, the quadrature variances of the
cavity radiation reduces to
\begin{align}\label{p51}\Delta c_{\pm}^{2}& = 1+
\langle\alpha^{*}(t)\alpha(t)\rangle +
\langle\beta^{*}(t)\beta(t)\rangle\notag\\&\pm\left[\langle\alpha^{*}(t)\beta^{*}(t)\rangle
 + \langle\alpha(t)\beta(t)\rangle\right].\end{align}
In order to understand the dependence of the evolution of the two-mode squeezing on the linear gain coefficient, amplitude of the external driving radiation, the rate of dephasing, and the phase fluctuation in depth,  case by case approach is preferred. Throughout this work, two-dimensional plots are chosen instead of the three-dimensional surfaces for the sake of clarity. It is worth noting that the main aim is geared to show the nature of the evolution so that comparison can be easily made notwithstanding the temptation that the maximum obtainable degree of squeezing can be optimized by adjusting the involved parameters.

\subsection{ Dynamics of the two-mode squeezing}

Let us first take the case when there is no phase fluctuation ($\theta=0$) and the atomic damping rate via every channel is the same ($\gamma=\Gamma$). In the following, the plots of the variance of the quadrature operators versus time is presented using Eqs. \eqref{p32}, \eqref{p33}, \eqref{p34}, \eqref{p45}, \eqref{p46}, \eqref{p47}, \eqref{p49}, \eqref{p50}, and \eqref{p51}.

\begin{figure}[hbt]
\centerline{\includegraphics [height=6.5cm,angle=0]{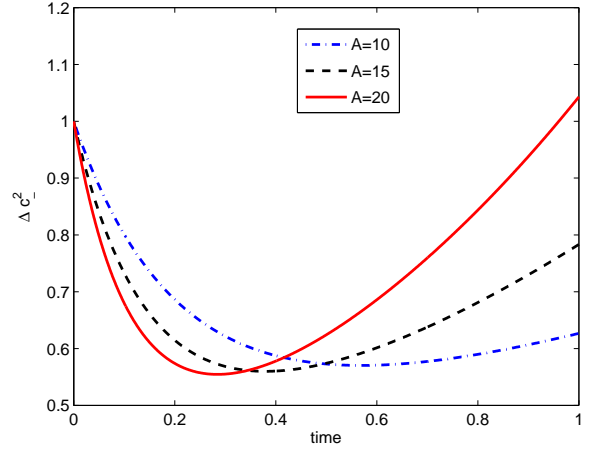}}
\caption {\label{fig1} Plots of the quadrature variance ($\Delta c_{-}^{2}$) of the cavity radiation for $\kappa=0.5$, $\gamma=\Gamma$, $\theta=0$, $\Omega=0.5\gamma$, and different values of $A$.}\end{figure}

\begin{figure}[hbt]
\centerline{\includegraphics [height=6.5cm,angle=0]{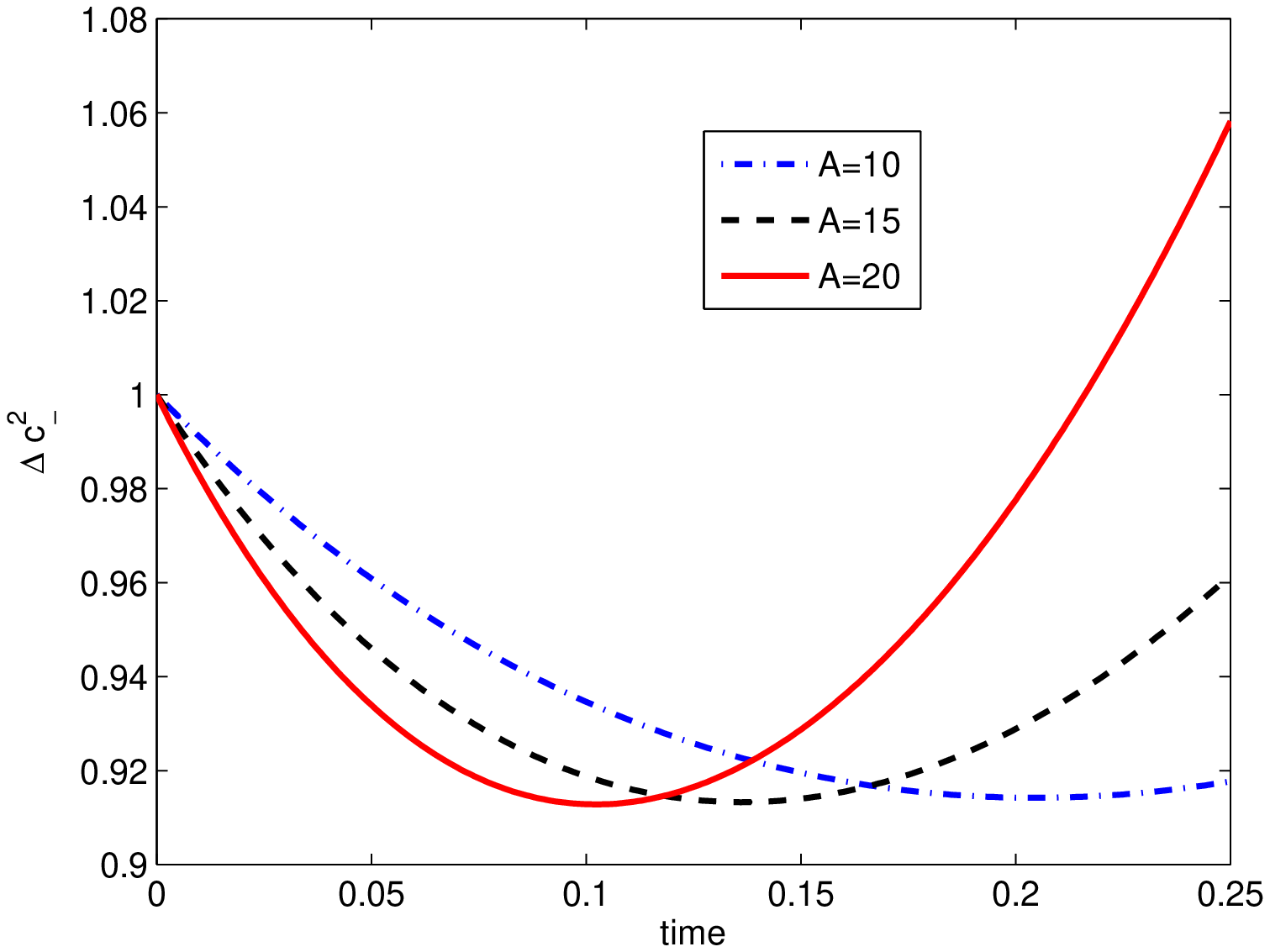}}
\caption {\label{fig3} Plots of the quadrature variance ($\Delta c_{-}^{2}$) of the cavity radiation for $\kappa=0.5$, $\gamma=\Gamma$, $\theta=0$, $\Omega=2.5\gamma$, and different values of $A$.}  \end{figure}

\begin{figure}[hbt]
\centerline{\includegraphics [height=6.5cm,angle=0]{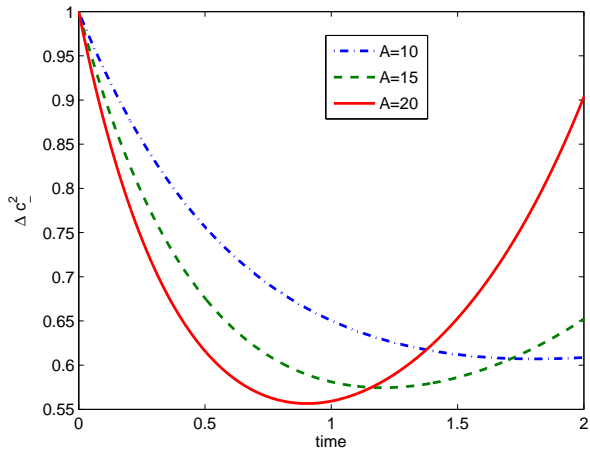}}
\caption {\label{fig2} Plots of the quadrature variance ($\Delta c_{-}^{2}$) of the cavity radiation for $\kappa=0.5$, $\gamma=\Gamma$, $\theta=0$, $\Omega=10\gamma$, and different values of $A$.}\end{figure}

It is clearly shown in Figs. \ref{fig1}, \ref{fig3}, and \ref{fig2} that the cavity radiation exhibits two-mode squeezing starting from the inception of the injection of the atoms. It is not difficult to notice that, in the early stages of the operation, the degree of squeezing rapidly increases with time irrespective of the values of the linear gain coefficient and amplitude of the external pumping radiation. This can be viewed as the injection of more atoms into the cavity leads to the generation of more correlated photons. This result in other words indicates that in the early stages of the amplification, the nonclassical properties of the radiation rapidly increase with time. However, as time passes by the thermal fluctuations arising due to heating begins to consume the established nonclassical features and hence the squeezing starts to wear away.  

The values of $\Omega/\gamma$ are chosen based on the earlier studies where the two-mode squeezing for similar case is found to be strong at $\Omega=0.5\gamma$, very poor when $\Omega=2.5\gamma$, and some what intermediate  when $\Omega=10\gamma$ at steady state \cite{jpb41055503}. In light of this, critical comparison of results shown in Figs. \ref{fig1}, \ref{fig3}, and \ref{fig2} asserts that the degree of squeezing is better for $\Omega=0.5\gamma$ in the early stages of the evolution. Quiet remarkably, contrary to steady state case, it is possible to witness squeezing for $\Omega=2.5\gamma$, although the degree of squeezing is still small when compared to the other two. Extrapolating this representative result evinces that it is possible to obtain squeezed light from this system for all values of the amplitudes of the external driving radiation in the early stages of the amplification process. It is also worth noting that the time at which a maximum squeezing can be manifested depends on the linear gain coefficient and the amplitude of the driving radiation. Critical scrutiny of Figs. \ref{fig1}, \ref{fig3}, and \ref{fig2} reveals that for similar values of $A$, the time at which the degree of squeezing would be maximum is the smallest when $\Omega=2.5\gamma$ and largest when $\Omega=10\gamma$. Most probably this can be related to the absence of squeezing when $\Omega=2.5\gamma$ at steady state. The other issue that may worth mentioning is the enhancement of the degree of two-mode squeezing with the linear gain coefficient at early stage of the operation and then the situation changes after maximum squeezing has been reached. This result also suports the explanation provided in relation to the thermal fluctuations earlier.

\subsection{Effect of the phase fluctuation on the dynamics of the two-mode squeezing}

The dependence of the evolution of the two-mode squeezing on the phase fluctuation for different values of the amplitude of the external driving radiation and when $\gamma=\Gamma$ is studied. $A=10$ is chosen for no particular reason.

\begin{figure}[hbt]
\centerline{\includegraphics [height=6.5cm,angle=0]{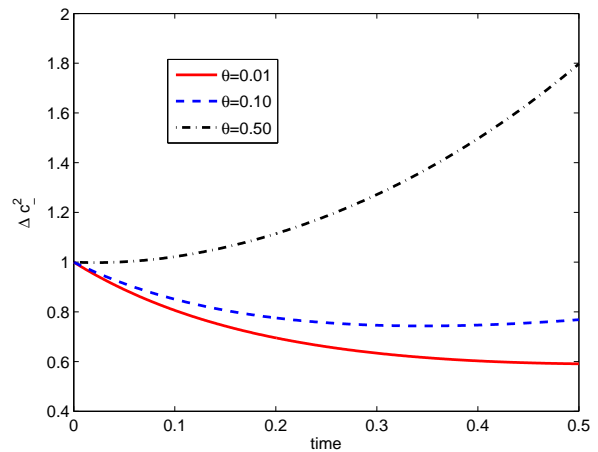}}
\caption {\label{fig4} Plots of the quadrature variance ($\Delta c_{-}^{2}$) of the cavity radiation for $\kappa=0.5$, $\gamma=\Gamma$, $A=10$, $\Omega=0.5\gamma$, and different values of $\theta$.} \end{figure}

\begin{figure}[hbt]
\centerline{\includegraphics [height=6.5cm,angle=0]{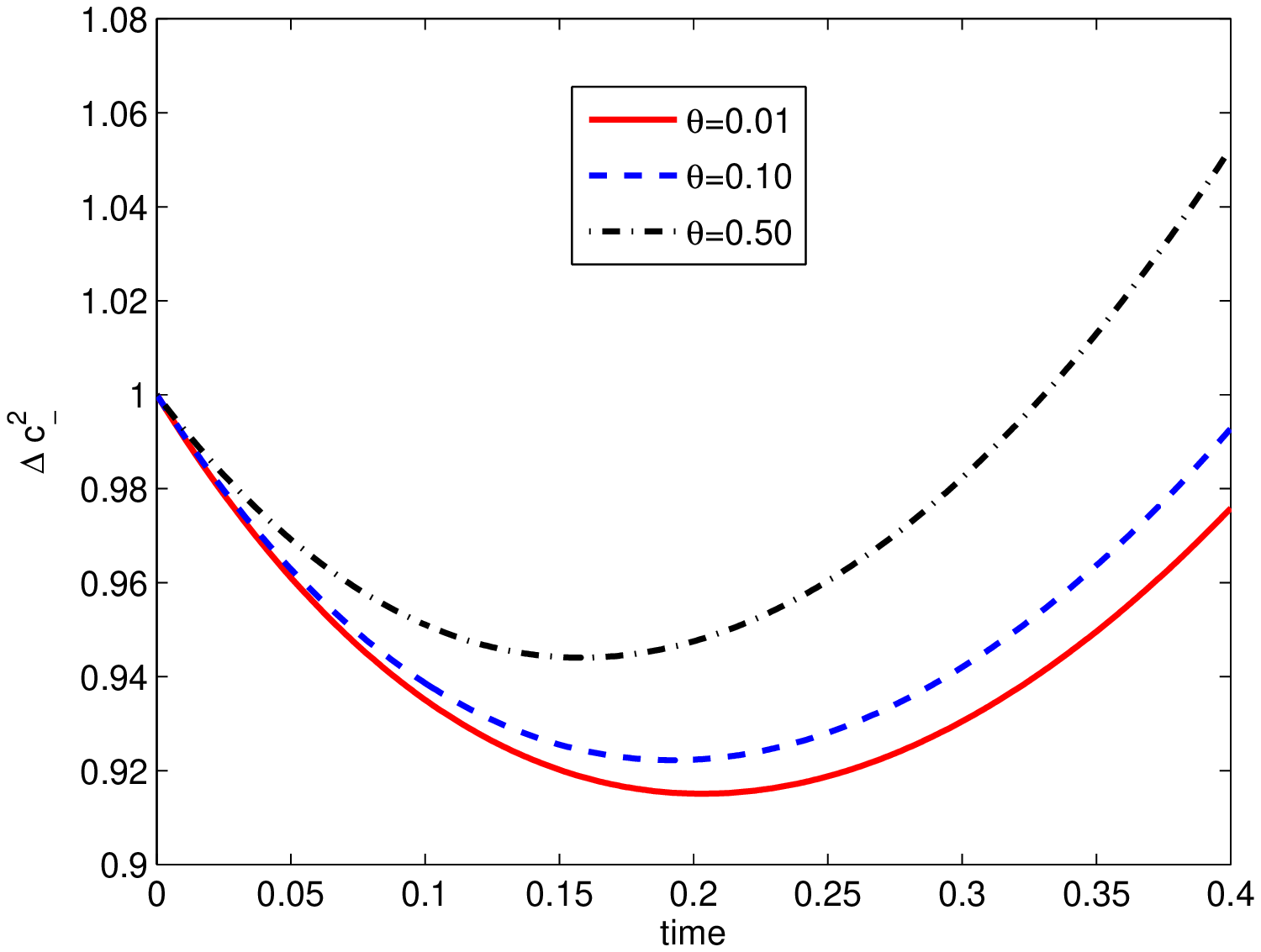}}
\caption {\label{fig5} Plots of the quadrature variance ($\Delta c_{-}^{2}$) of the cavity radiation for $\kappa=0.5$, $\gamma=\Gamma$, $A=10$, $\Omega=2.5\gamma$, and different values of $\theta$.} \end{figure}

\begin{figure}[hbt]
\centerline{\includegraphics [height=6.5cm,angle=0]{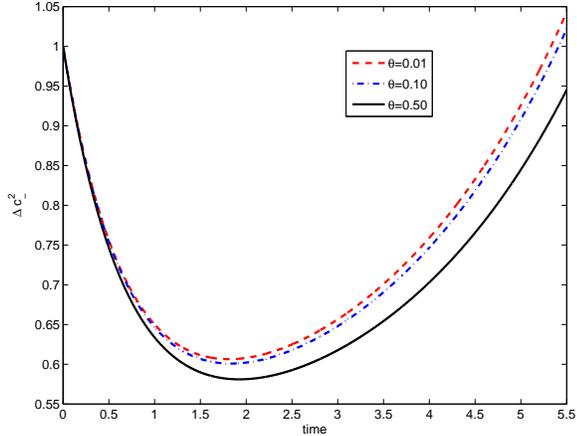}}
\caption {\label{fig6} Plots of the quadrature variance ($\Delta c_{-}^{2}$) of the cavity radiation for $\kappa=0.5$, $\gamma=\Gamma$, $A=10$, $\Omega=10\gamma$, and different values of $\theta$.} \end{figure}

It is not difficult to see from Figs. \ref{fig4}, \ref{fig5}, and \ref{fig6} that the two-mode squeezing increases with time in the early stages of the operation and wears away as time progresses for the reasons 	already explained. It is clearly shown in Figs. \ref{fig4} and \ref{fig5} that the degree of squeezing decreases with the enhancement of the phase fluctuation. Contrary to this, in Fig. \ref{fig6}, the degree of squeezing is found to increase with the phase fluctuation. The same phenomenon has been repeatedly observed when $\Omega$ is taken to be much greater than $\gamma$. Critical comparison of the results shown in Figs. \ref{fig2} and \ref{fig6}, for $A=10$, indicates that the two-mode squeezing is better when the atoms are initially prepared in a partial coherent superposition. This result entails that the squeezing that could have been lost due to phase fluctuation can be compensated for better by the external driving mechanism and vice versa via creating indirect roots for spontaneous atomic emission.  The other noteworthy aspect of the phase fluctuation, in the strong driving limit, is the relaxation of the time at which the maximum squeezing is manifested and the time for which the radiation remains in the squeezed state which  might be useful in practical situation in connection with the smallness of the actual time.

\subsection{Effect of dephasing on the dynamics of the two-mode squeezing}

In the following, upon fixing $A=10$ and varying the values of $\gamma/\Gamma$  the effect of dephasing on the evolution of the two-mode squeezing is analyzed. Here the rate at which the coherent superposition decays  is assumed to be greater than the other atomic decay rates. This assumption appears to be more realistic than assuming the two variant of atomic dampings as equal, since the induced atomic coherent superposition can be lost via quantum processes like vacuum fluctuations and atomic broadening without directly involving atomic transitions. Basically all phenomena related to atomic broadening, whereby the coherence between the lower and upper energy levels can be lost without necessarily one of them decaying to other levels, can lead to enhanced dephasing. Apparently, the effects of dephasing on the evolution of the two-mode squeezing is believed to be an interesting issue by its own right.

\begin{figure}[hbt]
 \centerline{\includegraphics [height=6.5cm,angle=0]{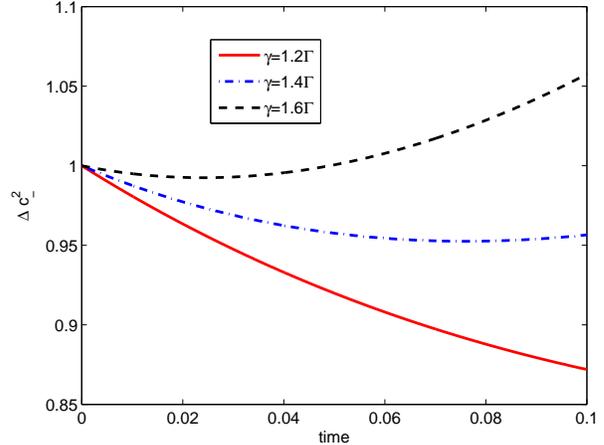}}
 \caption {\label{fig7} Plots of the quadrature variance ($\Delta c_{-}^{2}$) of the cavity radiation for $\kappa=0.5$, $\theta=0$, $A=10$, $\Omega=0.5\gamma$, and different values of $\gamma/\Gamma$.}
 \end{figure}

\begin{figure}[hbt]
 \centerline{\includegraphics [height=6.5cm,angle=0]{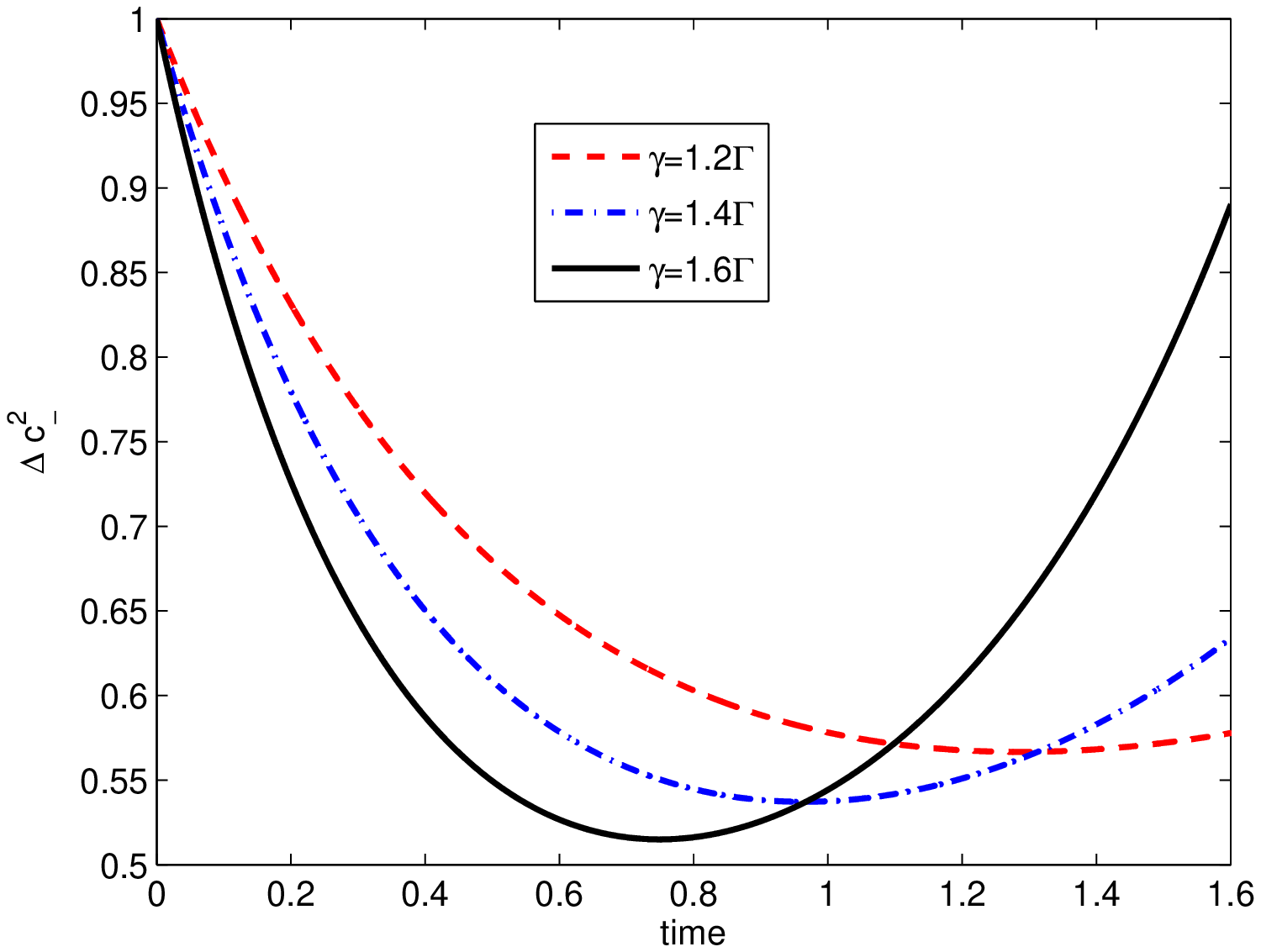}}
 \caption {\label{fig8} Plots of the quadrature variance ($\Delta c_{-}^{2}$) of the cavity radiation for $\kappa=0.5$, $\theta=0$, $A=10$, $\Omega=10\gamma$, and different values of $\gamma/\Gamma$.}
 \end{figure}

\begin{figure}[hbt]
 \centerline{\includegraphics [height=6.5cm,angle=0]{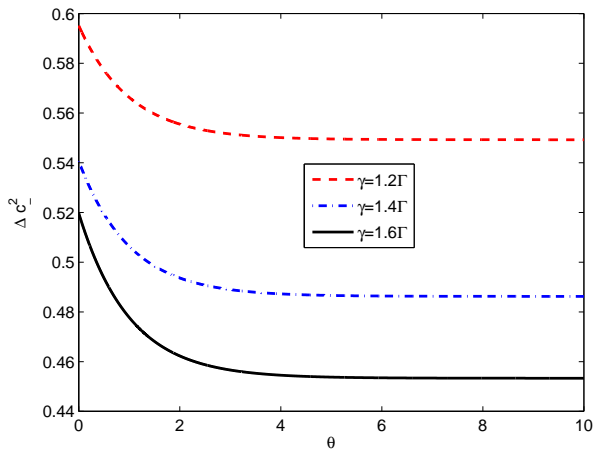}}
 \caption {\label{fig9} Plots of the quadrature variance ($\Delta c_{-}^{2}$) of the cavity radiation for $\kappa=0.5$, $t=0.85$, $A=10$, $\Omega=10\gamma$, and different values of $\gamma/\Gamma$.}
 \end{figure}

It is possible to infer from Fig. \ref{fig7} that the degree of two-mode squeezing decreases with increasing rate of dephasing for smaller amplitudes of the driving radiation. Similar to earlier cases the degree of two-mode squeezing increases with time at earlier stages and then decreases with time. It is worth to note that in addition to the squeezing the time at which the maximum squeezing obtained is also decreases with dephasing. Therefore, the practical advantage of dephasing in weak driving limit is far from encouraging. In the case that the phase can be locked via for instance coherent population trapping through two-step external pumping, that means when $\gamma<\Gamma$, the two-mode squeezing significantly increases with decreasing $\gamma/\Gamma$. A similar result has been recently reported for quadrature entanglement \cite{pra79063815}. Contrary to this, as clearly shown in Figs. \ref{fig8} and \ref{fig9}, the degree of two-mode squeezing increases with the rate of dephasing when the atoms are externally pumped with a strong radiation. It is straightforward to see that in Fig. \ref{fig6} the phase fluctuation is varied when $\gamma=\Gamma$, but in Fig. \ref{fig8} the dephasing is varied when $\theta=0$. In these cases, the degree of squeezing is found to increase with phase fluctuation and dephasing. On the other hand, as clearly shown in Fig. \ref{fig9}, both dephasing and phase fluctuation are made to vary. It turns out that the degree of two-mode squeezing increases with both. This result testifies that the assumptions that have been made for limiting the mathematical rigor have overshadowed the obtainable nonclassicality of the radiation. What surprises most is these assumptions have been thought to increase the same which of course is true in weak driving limit or when there is no driving at all. 

\section{Mean number of photon pairs}

In relation to the brightness of the squeezed light to be generated, it is necessary to evaluate the mean photon number which is directly associated with the intensity of the radiation. In order to attest the potential of the system under consideration as a reliable source of a two-mode squeezing, the evolution of the photon number for the cavity radiation would be studied. Genrally, the mean number of photon pairs describing the two-mode cavity radiation can be defined as
\begin{align}\label{p52}\bar{N}=\langle\hat{c}^{\dagger}(t)\hat{c}(t)\rangle,\end{align}
where $\hat{c}(t)$ is the annihilation operator given by
\eqref{p35}. 

On the basis that the operators in Eq. \eqref{p52} are already
put in the normal order, it is possible to rewrite it in terms of
 c-number variables associated with the normal ordering as
\begin{align}\label{en45}\bar{N}={1\over2}\big[\langle\alpha^{*}(t)\alpha(t)\rangle
+\langle\beta^{*}(t)\beta(t)\rangle\big].\end{align} Since $\langle\alpha^{*}(t)\alpha(t)\rangle$ and $\langle\beta^{*}(t)\beta(t)\rangle$ represent the mean photon number in mode $a$ and mode $b$, $\bar{N}$ is interpreted as mean number of photon pairs. The actual intensity of the radiation in the cavity can be taken as twice of $\bar{N}$ without essential lose of generality. In the following, to save time and space, only certain representative points are selected. The parameters in the analysis are fixed to be the same as the corresponding two-mode squeezing except in certain cases where  some changes are made for the purpose of clarity.

\begin{figure}[hbt]
\centerline{\includegraphics [height=6.5cm,angle=0]{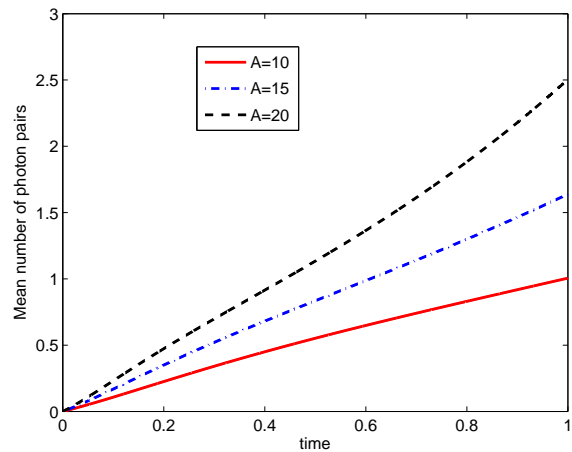}}
\caption {\label{fig10} Plots of the mean number of photon pairs ($\bar{N}$) of the cavity radiation for $\kappa=0.5$, $\gamma=\Gamma$, $\theta=0$, $\Omega=0.5\gamma$, and different values of $A$.}\end{figure}

\begin{figure}[hbt]
\centerline{\includegraphics [height=6.5cm,angle=0]{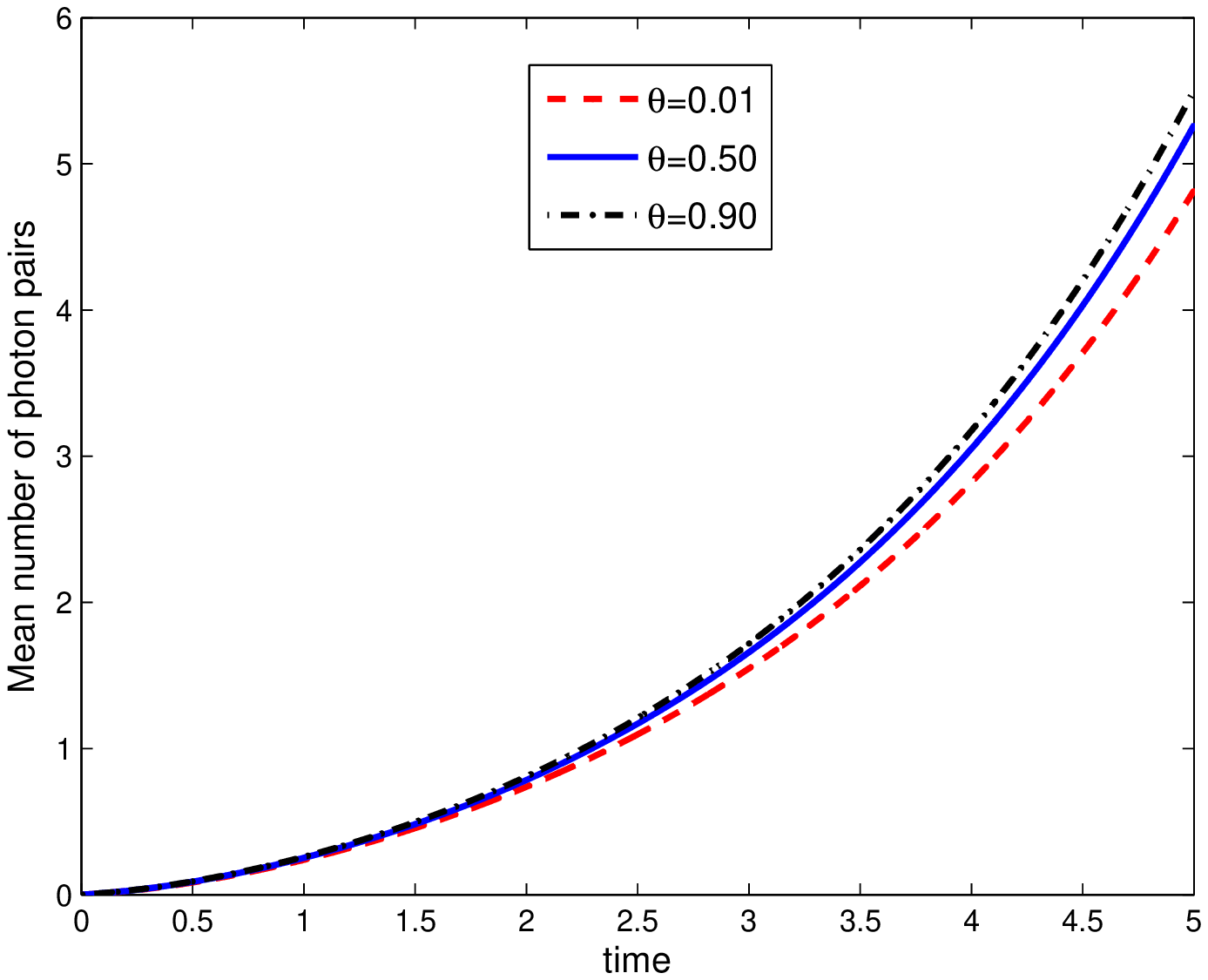}}
\caption {\label{fig11} Plots of the mean number of photon pairs ($\bar{N}$) of the cavity radiation for $\kappa=0.5$, $\gamma=\Gamma$, $A=10$, $\Omega=10\gamma$, and different values of $\theta$.}\end{figure}

\begin{figure}[hbt]
\centerline{\includegraphics [height=6cm,angle=0]{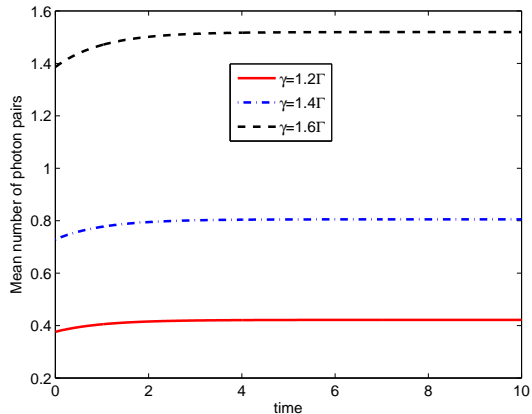}}
\caption {\label{fig12} Plots of the mean number of photon pairs ($\bar{N}$) of the cavity radiation for $\kappa=0.5$, $t=0.85$, $A=10$, $\Omega=10\gamma$, and different values of $\gamma/\Gamma$.}
 \end{figure}

It is not difficult to see from Figs. \ref{fig10}, \ref{fig11}, and \ref{fig12} that the intensity of the radiation increases with time irrespective of the changes in the parameters under consideration. Unlike the two-mode squeezing, this trend prevails even at larger values of time. This outcome mainly evinces that the more the atoms are injected into the cavity, the more photons are available in the cavity. The thermal fluctuations arising from heating of the cavity that has been found to consume the nonclassicality has no effect on the number of the photons to be generated. In line with this, it is possible to envision that this system can generate sufficiently bright light at steady state as already predicted in earlier communications \cite{jpb41055503}. It is also vividly presented in Fig. \ref{fig10} that the intensity of the radiation increases with linear gain coefficient in low driving limit ($\Omega<\gamma$). Since the atoms are initially prepared partially to be 50\% in the upper energy level, the more atoms are injected into the cavity implies more radiation would be emitted when the amplitude of the external radiation is too small to reverse the trend. Moreover, as clearly shown in Figs. \ref{fig11} and \ref{fig12}, the intensity of the radiation increases with the rate of dephasing and degree of phase fluctuation for selected parameters. 

\section{Discussion and conclusion}

In this contribution, analysis of the time evolution of the two-mode squeezing and intensity of the cavity radiation of a coherently pumped two-photon correlated emission laser is presented. The effect of dephasing resulting from the involved quantum phenomena like vacuum fluctuations and atomic broadening along with the phase fluctuation corresponding to incapability of preparing the atoms initially in a perfect 50:50 probability to be in the upper and lower energy levels on the dynamics of the squeezing are the main issues. In order to see the contribution of the driving radiation in depth, three representative cases are selected: the regime of weak driving limit ($\Omega=0.5\gamma$),  the case where the squeezing is expected to be minimum ($\Omega=2.5\gamma$), and  a strong driving limit ($\Omega=10\gamma$). In order to enhance clarity of exposition, the presentation is divided into three main categories. In the first place, the evolution of the two-mode squeezing due to changes in the linear gain coefficient and amplitude of the driving radiation is treated. Secondly, effect  of the phase fluctuation along with the variations in the amplitude of the external driving radiation on the evolution of the two-mode squeezing and intensity of the radiation is investigated. Lastly, effect of dephasing in relation to modifications in amplitude and phase fluctuation on the same is addressed. Although this study is limited to only few choices of the involved parameters, the obtained outcomes entail that the approach adopted can lead to a more tractable results than the approximated numerical analysis can offer.

It turns out that the degree of two-mode squeezing and the intensity of the generated radiation increase with time at the early stages of the amplification irrespective of the alterations in the parameters under consideration. However, this tendency changes as time progresses specially in case of the squeezing. It may worth mentioning that a similar conclusion has been drawn earlier under various conditions \cite{oc283781,pra444688,pra72022305,pra79013831}. On the other hand, the degree of two-mode squeezing begins to rapidly decrease with time whereas the intensity of the radiation remains to increase in later stages of the operation. It is not difficult to envision that with increment of time more atoms under went spontaneous emission while traversing across the cavity. Undoubtedly, the more there are emitted photons available in the cavity, the more would be the temprature of the cavity. Due to this heating, the atoms on the wall of the cavity starts to vigorously vibrate which causes additional thermal fluctuations. It is this thermal fluctuation that consumes the correlation to be established between the modes and consequently leads to inhibition of the nonclassical features. Since the thermal fluctuations do not have a direct impact on the number of emitted photons, the intensity of the radiation continues increasing with time as long as the atoms are injected into the cavity. Even though it is not evident in this study, it has been reported elsewhere that the generated radiation regains its nonclassical behavior at steady state since the coherence due to correlation of emitted photons slowly overcomes the effect of the thermal fluctuation through time \cite{pra74043816,jpb41055503}.

The dependence of the degree of two-mode squeezing on the dephasing and phase fluctuation is found to be different in the weak and strong driving limits. In the weak driving limit, the two-mode squeezing decreases with both as usually expected. However, in the strong driving limit,  the two-mode squeezing increases with both dephasing and phase fluctuation. Evaluating the situations from the root betoken that, in the strong driving limit, the coherence induced by external radiation is too strong to allow spontaneous emission in the cascading process. The evidence may be traced to a significant decrement in the mean photon number with the amplitude of the driving radiation in a similar case at steady state \cite{jpb41145501}. Hence dephasing and phase fluctuation characteristically weaken this coherence whereby provide an indirect electron path way which enhances the emission of correlated photons.

Based on the assumption that there is no interaction between the injected atoms, that is clearly manifested in Eq. \eqref{p48}, the relation between the two-mode squeezing and the entanglement measure due to the criterion following from Duan {\it{et al.}} \cite{prl842722} for general nondegenerate three-level laser has been established \cite{pra77013815,jpb41145501}. In view of these reports, the results obtained for the degree of two-mode squeezing are equally applicable for the quadrature entanglement. With this understanding, it is possible to conclude that the technical inability of preparing the atoms in a perfect coherent superposition and the failure to control the destruction of coherence by quantum fluctuations provide a means for enhancing the quantum features including entanglement. If such unexpected outcomes can be properly harnessed, hopefully they support the pursuit for generating strong light with robust nonclassical features. Moreover,  the possibility of increasing the time at which maximum squeezing is observed and the time in which the light remains in the squeezed state by manipulating the phase fluctuation can be utilized in the practical realization of this system. Furthermore, the increment of the intensity of the radiation with time complements the advantages of this system as a potential candidate for the generation of the light with strong nonclassical features. Since the assumptions that have been made for some particular interest (which could have posed a sever restriction) are lifted, it is hoped that the system under consideration can be tested in a more relaxed manner with anticipated far better outcomes. With no doubt, this would be a positive signature in the pursuit of generating strong nonclassical light from this system.

\section*{Acknowledgments}

I thank the Max Planck Institute for Physics of the Complex Systems for allowing me to visit them and use their facility in carrying out this research and Dilla University for granting the leave of absence.


\begin{thebibliography}{1}
\bibitem{pra74043816} S. Tesfa, Phys. Rev. A {\bf{74}},   043816 (2006).
\bibitem{pra49481} J. Anwar and M. S. Zubairy, Phys. Rev. A {\bf{49}}, 481 (1994).
\bibitem{jpbamos} S. Tesfa, J. Phys. B: At. Mol. Opt. Phys. {\bf{40}},   2373 (2007).
\bibitem{prl601832} M. O. Scully, K. Wodkiewicz, M. S. Zubairy, J. Bergou, N. Lu, and M. Meyer ter Vehn, Phys. Rev. Lett. {\bf{60}}, 1832 (1988).
\bibitem{pra72022305} H. T. Han, S. Y. Zhu, and M. S. Zubairy, Phys. Rev. A {\bf{72}}, 022305 (2005).
\bibitem{jpb41145501} S. Tesfa, J. Phys. B: At. Mol. Opt. Phys. {\bf{41}}, 145501 (2008).
\bibitem{prl94023601} H. Xiong, M. O. Scully, and M. S. Zubairy, Phys. Rev. Lett. {\bf{94}}, 023601 (2005).
\bibitem{pra75033816} M. Kiffner, M. S. Zubairy, J. Evers, and C. H. Keitel, Phys. Rev. A {\bf{75}}, 033816 (2007).
\bibitem{pra415179} N. A. Ansari, J. Gea-Banacloche, and M. S. Zubairy, Phys. Rev. A {\bf{41}}, 5179 (1990).
\bibitem{pra461560} N. A. Ansari, Phys. Rev. A {\bf{46}}, 1560 (1992).
\bibitem{pra484686} N. A. Ansari, Phys. Rev. A {\bf{48}},   4686 (1993).
\bibitem{sint} S. Tesfa, arXiv:1009.1568.
\bibitem{pra79013831} S. Qamar, M. Al-Amri, and M. S. Zubairy, Phys. Rev. A {\bf{79}}, 013831 (2009).
\bibitem{pra444688} M. Majeed and M. S. Zubairy, Phys. Rev. A {\bf{44}}, 4688 (1991).
\bibitem{oc283781} Sajid Qamar, Shahid Qamar, and M. S. Zubairy, Opt. Commun. {\bf{283}}, 781 (2010).
\bibitem{pra79063815} S. Tesfa, Phys. Rev. A {\bf{79}}, 063815 (2009).
\bibitem{pra77062308} S. Qamar, F. Ghafoor, M. Hillery, and M. S. Zubairy, Phys. Rev. A {\bf{77}}, 062308 (2008).
\bibitem{jpb41055503} S. Tesfa, J. Phys. B: At. Mol. Opt. Phys. {\bf{41}}, 055503 (2008).
\bibitem{pra75062305} S. Qamar, H. Xiong, and M. S. Zubairy, Phys. Rev. A {\bf{75}}, 062305 (2007).
\bibitem{pra79033810} S. Tesfa, Phys. Rev. A {\bf{79}}, 033810 (2009).
\bibitem{scully} M. O. Scully and M. S. Zubairy, {\it{Quantum Optics}} (Cambridge University Press, Cambridge, 1997).
\bibitem{lou} W. H. Louisell, {\it{Quantum statistical properties of radiation}} (Wiley, Newyork, 1973).
\bibitem{pra77013815} S. Tesfa, Phys. Rev. A {\bf{77}}, 013815 (2008).
\bibitem{method}S. M. Barnett and P. M. Badmore, {\it{Methods in theoretical quantum optics}} (Oxford University Press, New York, 1997).
\bibitem{prl842722} L. M. Duan, G. Giedke, J, I, Cirac, and P. Zoller, Phys. Rev. Lett {\bf{84}},   2722 (2000).



\end{thebibliography}
\end{document}